\documentclass[aip,jmp,amsmath,amssymb,reprint]{revtex4-1}

\usepackage{graphicx}
\usepackage{dcolumn}
\usepackage{bm}

\begin{document}

\preprint{AIP/123-QED}

\title[On the Efficiency of Underground Systems in Large Cities]{On the Efficiency of Underground Systems in Large Cities}

\author{Luciano da Fontoura Costa}
\email{ldfcosta@gmail.com}
\altaffiliation{Also at National Institute of Science and Technology for Complex Systems, Brazil.}

\author{Bruno Augusto Nassif Traven\c{c}olo}

\author{Matheus Palhares Viana}
\affiliation{Institute of Physics\\University of S\~ao Paulo, S\~ao Carlos, S\~ao Paulo, Brazil.}

\author{Emanuele Strano}
\affiliation{Urban Design Studies Unit, Department of Architecture\\University of Strathclyde, Glasgow, UK.}

\date{\today}

\begin{abstract}
We report an analysis of the accessibility between different locations
in big cities, which is illustrated with respect to London and Paris.
The effects of the respective underground systems in facilitating more
uniform access to diverse places are also quantified and
investigated. It is shown that London and Paris have markedly
different patterns of accessibility, as a consequence of the number of
bridges and large parks of London, and that in both cases the
respective underground systems imply in general increase of
accessibility.
\end{abstract}



\maketitle

\begin{figure}[ht]
\begin{center}
    \includegraphics[scale=0.5]{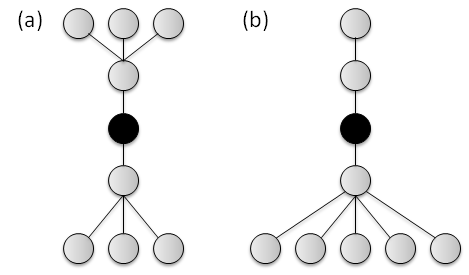}
\end{center}
\caption{ Diversity entropy and balanced access. Considering
self-avoiding random walks, in (a) the black node can access six
other nodes within two steps with equal probability, while in (b)
different probabilities are found when accessing the nodes after two
steps. These properties are captured by the diversity entropy, which
is lower for the case (b) ($E_2=0.72$) when compared with (a) ($E_2
=0.86$).}\label{fig:exdv}
\end{figure}

Underground systems have a critical rule in metropolitan areas,
affecting urban economic development, people transportation systems,
car dependence with related pollution problems, as well as the urban
form. Therefore, in the global debate about urban sustainability,
underground and other transportation systems were and still are big
and controversial issues, to the extent that the quantification of
their impact represents a challenge for city planners and local
administrations \cite{Newman1999,Marshall2001,Banister2007}. In this
letter, we report an analysis of the accessibilities and the effects
of the underground systems in the central area of London and Paris
street network (area of 13 km$^2$).  In order to do so, we first
transform the streets of these cities into complex networks
\cite{Barabasi2002Survey,Barabasi2007,Costa2007Survey} so that every
confluence of three or more streets becomes a node, while the streets
themselves correspond to the links, and then calculate the diversity
entropy of each node. The latter measurement is particularly useful to
quantify the potential of each node in accessing in a balanced and
homogeneous manner other nodes at specific scales. The mathematical
definition of the diversity entropy of the node $i$ is given as
\cite{Travencolo2008,Travencolo2008-2,Travencolo2009}:

\begin{equation}
  E_h(i) = -\frac{1}{log(N-1)} \sum_{j=1}^{N}
          P_h(j,i)log(P_h(j,i))
  \label{eq:diversity}
\end{equation}

for $P_h(j,i)\neq 0$, where $P_h(j,i)$ corresponds to the
probability that an agent performing a self-avoiding random walk
reaches the node $j$ after $h$ steps departing from $i$.  The
higher the diversity entropy for a given $h$, the more balanced
will be the access from node $i$ to the other nodes at that
distance.  Figure~\ref{fig:exdv} illustrates the concept of
diversity entropy for two simple graphs with different
interconnecting patterns.

The data of the central areas of Paris and London were extracted from
the OpenStreetMap website (http://openstreetmap.org) using the
software Merkaartor \cite{markurl}, which supplies information about
both the positions of streets and underground routes. The street
networks were coupled to their respective underground networks by
linking every underground station to the closest node (considering
Euclidean distance) of the street network. The final networks of Paris
and London resulted with 11699 (669 from underground) nodes and 6885
(346 from underground) nodes, respectively. The average degrees of
these networks are 3.02 for Paris and 2.73 for London.

The displacements of people through the cities were modeled in terms
of self-avoiding random walks, so as to avoid going back along the
routes. Two different types of walks were considered: respectively to
the network with and without the underground system. For the latter,
self-avoiding random walks were used. For the former case, i.e.\
including the underground system, several diffusive process starting
from each station node were used so that the sum of every diffusion
was considered as the weight of the nodes. The agents move over the
streets network following the gradient of the nodes weights. In this
approach the station nodes can be considered as local attractors of
agents and these agents are naturally guided to the closest
underground station. When an agent reaches a station it enters in the
metro and navigates through the underground by traditional random
walk. When it leaves the metro, the weights of the nodes are inverted,
so that the station works as a repeller and the agent continues his
self-avoid random walk in the nodes of the streets.

\begin{figure*}[ht]
\begin{center}
    \includegraphics[scale=0.5]{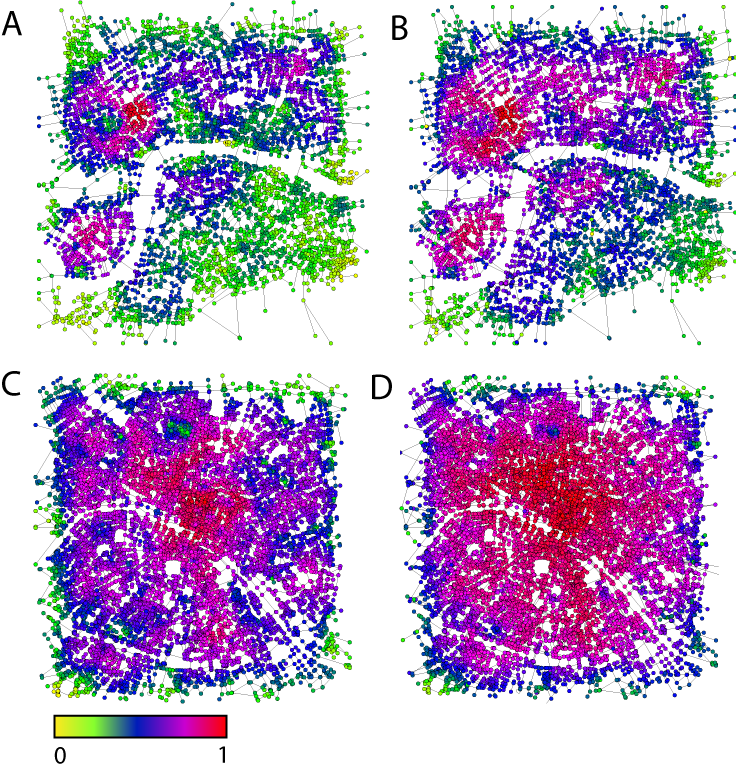}
\end{center}
\caption{The diversity values obtained for London without (A) and with
underground (B). Diversity of Paris without (C) and with underground
(D). In both cities, the incorporation of the respective
undergrounds led to higher and more uniform diversity
values.}\label{fig:DV}
\end{figure*}

\begin{figure}[ht]
\begin{center}
    \includegraphics[scale=0.6]{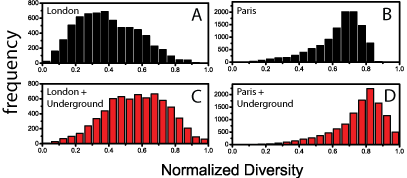}
\end{center}
\caption{ The histograms of diversity values for London without (A)
and with underground (C). The histograms for Paris without (B) and
with (D) the underground system.}\label{fig:hist}
\end{figure}

The diversity entropy was estimated for both cities with and without
the underground network and a total of 10,000 walks were performed for
each node, for every city. Each walk had length equal to the average
shortest path length of the respective network (i.e.\ 41 steps for
London and 37 for Paris steps). The obtained results are depicted in
Figure~\ref{fig:DV}, where the diversity entropy values are coded by
colors. The values of diversity obtained for the two cities are
organized as histograms in Figure~\ref{fig:hist}. Several interesting
insights can be inferred from this figure. Describing the
morphological structure of both cities, we have that Paris has more
uniform distribution of diversity entropy than London, meaning more
homogeneous access between generic locations, probably as a
consequence of its boulevards pattern. Second, as it can be clearly
appreciated from Figures~\ref{fig:DV}, the incorporation of the
respective underground systems clearly enhanced the diversity entropy
values and homogeneity in both cases. It is possible to note that the
efficiency of London street network is more dependent on the
underground system than in Paris, reflecting particularly well their
respective different mutual dependence on urban evolution.  It is also
interesting to note that the fact that the river Thames is much wider
than the Seine implied in the southern part of London being less
integrated into the remainder network than the counterpart situation
in Paris (the Seine has almost 2.5 times more bridges along central
Paris than the Thames within central London).  Even when restricted to
the northern regions, London still has less uniform accessibility,
which is possibly a consequence of the large parks found in the
central region of that city (Regent's and Hyde Parks).

The obtained results reinforces that the diversity entropy is a
particularly sensitive method for revealing relationships between
urban form and their underground systems. Hence there are many
applications of the reported methodology, especially in assisting
transportation, city planning, and urban form research.

\section*{Acknowledgments}

Luciano da F. Costa thanks FAPESP (05/00587-5) and CNPq (301303/06-1)
for sponsorship. B. A. N. Traven\c{c}olo is grateful to FAPESP
(07/02938-5) for financial support. Matheus P. Viana thanks FAPESP
(07/50882-9) for financial support.

\nocite{*}
\bibliography{subway}
\end{document}